# Unraveling the Molecular Structure of Lipid Nanoparticles through *in-silico* Self-Assembly for Rational Delivery Design

Xuan Bai*, Yu Lu, Tianhao Yu, Kangjie Lv, Cai Yao, Feng Shi, Andong Liu, Kai Wang*, Wenshou Wang, Chris Lai

METiS Technology (Hangzhou) Co., Ltd, Hangzhou, China, 310000

Emails: xbai@metispharma.com, baixuan@zju.edu.cn; kwang@metispharma.com.

**Abstract**

Lipid nanoparticles (LNPs) are a leading platform in the delivery of RNA-based therapeutics, playing a pivotal role in the clinical success of mRNA vaccines and other nucleic acid drugs. Their performance in RNA encapsulation and delivery is critically governed by the molecular structure of ionizable lipids and the overall formulation composition. However, mechanistic insight into how these factors govern LNP architecture and function remains limited, primarily owing to the challenges of capturing nanoscale assembly and organization using experimental techniques. Here, we employ coarse-grained molecular dynamics simulations to systematically investigate how ionizable lipid chemistry influences LNP self-assembly, internal organization, and surface properties. We further explore the effects of formulation ratios and pH-dependent deprotonation on both the internal structure and surface morphology of LNPs. Leveraging these insights, we demonstrate how *in silico* structural characteristics can inform the rational design of novel ionizable lipids and optimization of formulation ratios, supported with experimental validations. Our findings offer a molecular-level understanding of LNP assembly dynamics and architecture, thereby establishing a computational framework linking lipid chemistry and LNP formulation to the structure and performance of LNP, to advance the rational design of novel LNP delivery systems.

**Introduction**

The delivery of nucleic acid therapeutics has become a cornerstone of modern gene therapy and vaccine development.[1] Lipid nanoparticles (LNPs) have proven indispensable for transporting these fragile biomolecules, offering protection against enzymatic degradation, promoting cellular uptake, and enabling efficient intracellular release.[2-4] The rapid development and deployment of mRNA vaccines during the COVID-19 pandemic, notably those from Pfizer-BioNTech and Moderna, underscore the transformative potential of LNPs in clinical applications.[5, 6] By improving the pharmacokinetics, biodistribution, and stability of nucleic acids, LNPs have greatly expanded the scope and efficacy of precision medicine.[7, 8]

LNPs derive their function from their molecular architecture, which arises from the spontaneous self-assembly of four primary components: ionizable lipids, helper lipids, cholesterol, and PEGylated lipids[9, 10]. Each component plays a distinct yet synergistic role. Ionizable lipids are central to RNA encapsulation and endosome escape for cargo release: they adopt a positively charged, cone-shaped conformation at acidic pH, facilitating RNA encapsulation and endosomal escape, while becoming neutral at physiological pH to minimize cytotoxicity[10, 11]. Helper lipids (e.g., phospholipids) and cholesterol contribute to membrane stability and fluidity, whereas PEG-lipids form a hydrophilic shield that enhances colloidal stability and reduces protein adsorption[9, 12]. These components are combined at specific molar ratios to ensure optimal LNP structure and function. Consequently, the development of novel ionizable lipids and the optimization of formulation ratios have become focal points in LNP innovation.[13-16]

Despite substantial advances, understanding how ionizable lipid chemistry and formulation ratios collectively affect LNP structure and performance remains a key barrier to rational design. Advanced experimental techniques such as cryogenic transmission electron microscopy (cryo-TEM)[17, 18], small-angle X-ray scattering[19], and dynamic nuclear polarization-enhanced NMR[20, 21] have provided valuable structural insights, yet they are limited to resolve the dynamic processes underlying LNP self-assembly and molecular organization. Molecular dynamics (MD) simulations bridge this gap by offering molecular interactions over time, enabling direct observation of the stepwise construction of LNPs from their constituent components[22-25]. Particularly, coarse-grained (CG) MD simulations allow modeling over extended time and length scale that reaches tens of nanometers and microseconds, making them well-suited to study LNPs at physiologically relevant dimensions[26-30].

These simulations can reconstruct the self-assembly process and enable in silico screening of novel lipid structures and formulation combinations, significantly reducing the cost and time associated with extensive experimental screening. Crucially, they offer mechanistic insight into the structure–function relationships that underlie LNP efficacy.

In this study, we used CGMD simulations to capture the full self-assembly process of LNPs, starting from loosely distributed molecular components that spontaneously aggregate into small fragments and ultimately fuse into mature nanoparticles. By initializing the simulation with a tighter spatial packing of lipid components, we accelerated the assembly process that allowed us to simulate the formation of LNPs with diameters exceeding 50 nm, closely resembling the size of experimentally observed particles. We then systematically examined how formulation ratios and pH-dependent deprotonation affect both the internal architecture and surface characteristics of LNPs. Finally, we demonstrated how simulation-derived structural insights can be applied to guide the rational design of novel ionizable lipids and formulation strategies combining with experimental validation, offering a computational approach for accelerating the development of next-generation RNA delivery systems.

**Results**

**Stepwise self-assembly process of LNP and the effect of lipid chemistry**

Harnessing the amphiphilic nature of lipid components, LNPs can spontaneously self-assemble into enclosed structures capable of encapsulating RNA-based cargos[31]. The resulting internal organization is governed by the molecular structure of ionizable lipids and the molar ratio of each LNP component. Gaining mechanistic insight into the stepwise formation of LNPs from dispersed lipid monomers to mature nanoparticles is critical for elucidating the role of lipid chemistry in RNA encapsulation and the structural principles that underpin functional delivery.

To investigate this process, we first simulated the self-assembly of an MC3-containing LNP as an example. Lipid and RNA components were initially dispersed uniformly within a 50 nm aqueous simulation box (Fig. 1a). We selected short interfering RNA (siRNA) as the model nucleic acid cargo due to its relatively stable tertiary structure, in contrast to mRNA, whose conformation is ambiguous during dynamic assembly[32]. Over time, lipid and RNA molecules formed micellar or bilayer-like patches (<10 nm), which further aggregated into closed vesicular structures (>10 nm) containing both core- and surface-bound RNA (Fig. 1b). Higher-magnification views show that these vesicles undergo subsequent fusion, mediated by surface-localized RNA, resulting in larger, more stable LNPs. Reverse micelles that encapsulate RNA within their interiors gradually form as the process proceeds, while PEGylated lipids remain localized on the outer surface. These intermediate vesicles ultimately fuse to generate mature LNPs characterized by multiple internal reverse micelles and near a lipid monolayer envelope with PEGylation as seen in the cross-sectional view. A time-resolved visualization of the full self-assembly process of the MC3-based LNP is provided by Supplementary Video 1.

To explore how lipid chemistry impacts self-assembly, we conducted parallel simulations using two additional ionizable lipids, Lipid5 and ALC-0315, under identical conditions. While all three lipids followed a similar pathway from micellar aggregates to closed vesicles and ultimately fused LNPs (Fig. 1c-d), but the kinetics varied. Lipid5 and ALC-0315 exhibited significantly faster fusion compared to MC3 (Fig. 1e), attributed to their more pronounced cone-shaped molecular geometry and higher membrane fluidity, which facilitate lipid exchange and vesicle merging. These properties also yield smoother LNP surfaces than those formed by MC3. Despite similar assembly trajectories, the internal organizations differed markedly. ALC-0315 LNPs contained the highest number of the smallest reverse micelles, while MC3 LNPs had the fewest and largest reverse micelles, suggesting that ALC-0315 has the strongest propensity to form reverse micelles, followed by Lipid5, with MC3 being the weakest. This trend aligns with the relative delivery efficiencies of these LNPs[33], indicating that the strong ability to form reverse micelles may correlate positively with enhanced endosomal escape. We hypothesize that ionizable lipids capable of organizing into tightly curved reverse micelles may more readily facilitate the formation of the stalk intermediate that is a critical high-curvature step in the membrane fusion process with the endosomal membrane[34]. This interpretation is consistent with experimental findings demonstrating that inverse hexagonal phases can promote membrane fusion and enhance intracellular silencing efficacy relative to lamellar structures[17, 35]. Collectively, these results highlight the predictive potential of virtual screening based on internal LNP structure for evaluating delivery efficacy *in silico*.

The surface property of the LNPs was investigated by analyzing the surface hydrophobicity of the LNP surface (Fig. 1g). We noticed that the surfaces of these LNPs were all amphiphile, but the extent of hydrophobicity was dependent on lipid chemistry. MC3, with a single hydrophilic head group, yielded the least hydrophilic surface. Lipid5 and ALC-0315 share similar headgroup chemistries, but ALC-0315 contains four hydrocarbon tails versus three in Lipid5. This tail expansion introduced more steric hindrance, disrupting tight headgroup packing and further reducing surface hydrophilicity. These results underscore an inverse relationship between tail number and surface hydrophilicity, highlighting the role of lipid geometry in tuning interfacial characteristics. Finally, to evaluate LNP formation in a more computationally efficient setting, we conducted simulations in smaller systems with and without RNA cargos (Supplementary Fig. S1). These setups can also generate a complete but smaller LNP (Fig.1h) via self-assembly from dispersed components within shorter timescales and are particularly useful for high-throughput virtual screening of novel ionizable lipids and formulation conditions.

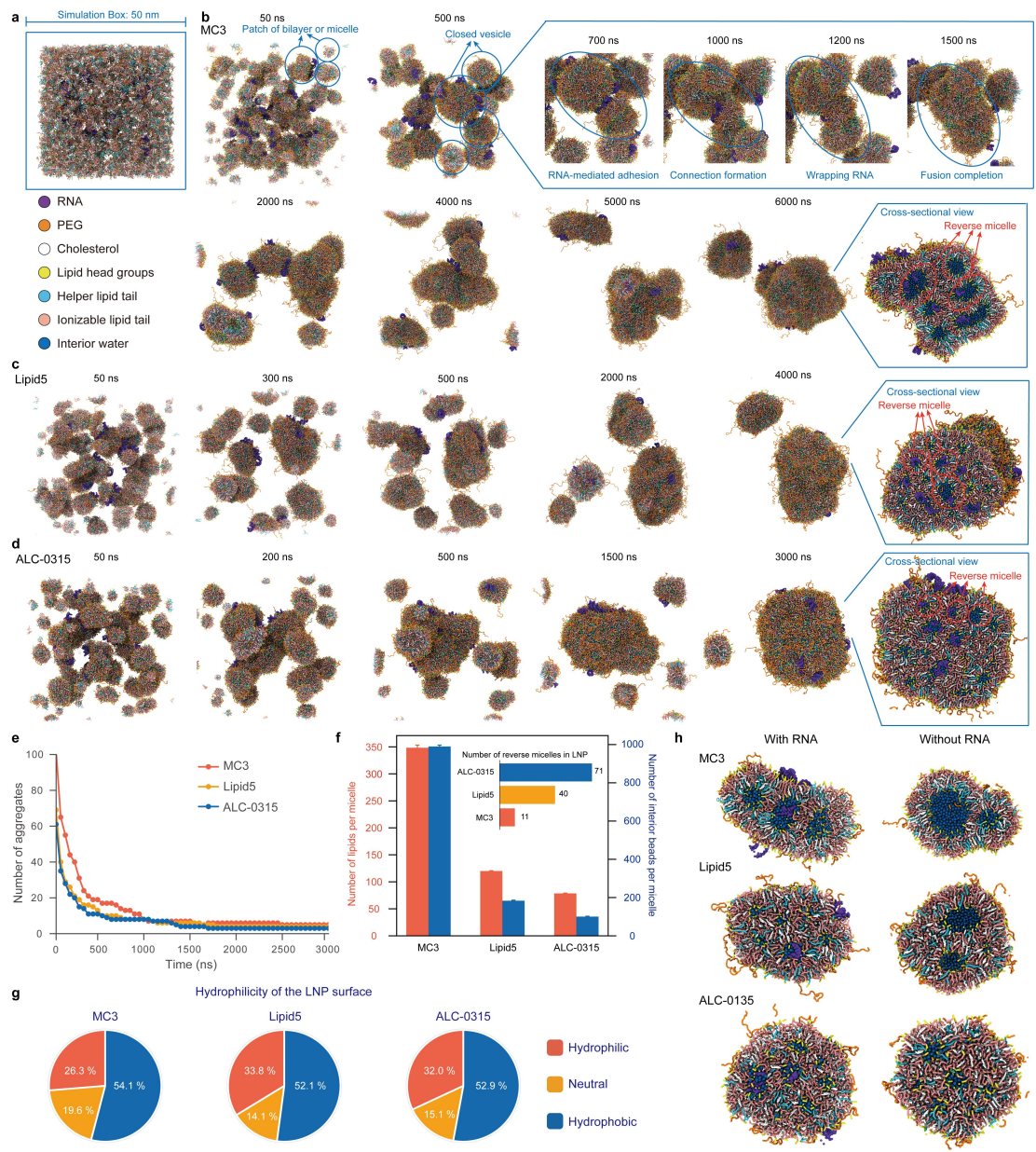

Fig. 1: Stepwise self-assembly of LNPs from dispersed components. a. Initial simulation setup, consisting of 5000 ionizable lipids, 4000 cholesterol molecules, 1000 phospholipids (DSPC), 150 PEG-lipids, 30 RNA strands, and nearly one million water beads (not shown). RNA is depicted in violet, PEG in PEG-lipids in orange, cholesterol in white, lipid head groups of helper lipids and ionizable lipids in yellow (highlighting the layered structure of the assembled LNP), lipid tails of helper and ionizable lipids in cyan and pink, and interior water in LNPs in blue. b-d. Time-resolved visualization of the self-assembly process for MC3 LNP (b), Lipid5 LNP (c), and ALC-0315 LNP (d). Insets in b illustrate key stages of MC3 LNP formation, including RNA-mediated adhesion, vesicle closure for RNA encapsulation,

and fusion completion. Cross-sectional views highlight the presence of reverse micelles in the final structures. e. Time evolution of the number of lipid aggregates during self-assembly, indicating difference in the fusion kinetics across ionizable lipid types. f. Quantification of lipid and interior content per reverse micelle in the final self-assembled LNP structures. The inset shows the number of reverse micelles in each LNP. g. Surface hydrophilicity of LNPs, represented as the relative fraction of hydrophilic, neutral, and hydrophobic regions at the interface of LNP and surrounding environments. h. Cross-sectional views of the final self-assembled LNP structures in a smaller simulation system, with and without RNA. The final LNP diameter is approximately 18 nm.

**Constructing full-scale LNP structures from self-assembly**

While the stepwise assembly method allows for dynamic tracking of LNP formation, it is computationally intensive and typically yields particles smaller than those observed experimentally. To overcome this limitation, we initiated simulations using densely packed lipids (Fig. 2a), after several hundred nanoseconds, enabling the construction of self-assembled LNPs over 50 nm in diameter with a water volume fraction of 24% in good agreement with earlier experimental work for the core phase in LNPs[19, 36] (Fig. 2b, Supplementary Fig. S2). This approach allowed us to assess the structural organizations and surface properties of LNPs at a more realistic scale. Compared to smaller LNPs (Fig. 1), these larger LNPs exhibited increased surface hydrophilicity and reduced hydrophobicity, which may enhance colloidal stability (Fig. 2c). These findings suggest a size-dependent relationship in which LNPs with more hydrophilic surfaces can remain stable at smaller diameters, whereas more hydrophobic LNPs may require larger sizes to achieve comparable stability.

However, this method restricts molecular mobility during the assembly process, making the final LNP structure highly sensitive to the initial molecular distribution. While increasing the initial lipid packing density did not significantly affect surface hydrophilicity (Supplementary Fig. S3), it reduced the internal water fraction from ~24% to ~15%, inducing a transition in MC3 LNP interiors from ellipsoidal to wormlike micellar structures[36, 37] (Fig. 2d and Supplementary Fig. S4). In contrast,

Lipid5 LNPs maintained spherical internal micelles, showing less sensitivity to changes in water content (Fig. 2d and Supplementary Fig. S4). Radial density profiles (Fig. 2e-h) further highlight the differences in internal lipid organization between the two formulations: MC3 LNPs exhibit more pronounced fluctuations in lipid component distribution than Lipid5 LNPs, consistent with a more stratified internal organization. The reduction in water content led to a smoother distribution of both lipids and water in the core, particularly in MC3 LNPs, resulting in more homogeneous interior structures. Additionally, lipid headgroups displayed two shallow density peaks at the surface of LNP under 24% water, indicative of coexisting bilayer and monolayer arrangements. These peaks disappeared when the interior water reduced to 15%, suggesting a structural transition toward a predominantly monolayer configuration.

This approach also enabled the construction of LNPs encapsulating larger RNA cargos (Fig. 2i). The inclusion of large RNA strands did not alter the surface hydrophilicity (Supplementary Fig. S5) and produced similar internal structures compared to LNPs containing small RNAs, provided the water fraction was constant (Fig. 2j). The visualization of the internal structure of the LNP showed that the larger RNAs appeared to be accommodated within interconnected aqueous compartments, reflecting their spatial requirements (Fig. 2k). Experimental studies have shown that mRNA-loaded LNPs typically require a higher internal water fraction than those encapsulating smaller RNA cargos such as siRNA[21]. In agreement with this, our simulations reveal that increased water content disrupts internal layering and promotes the formation of larger reverse micelles, thereby explaining the structural distinctions between siRNA- and mRNA-loaded LNPs[21, 38]. Additionally, this strategy can be extended to generate smaller LNPs (e.g., ~30 nm; Supplementary Fig. S6), facilitating high-throughput *in silico* screening of novel lipid formulations.

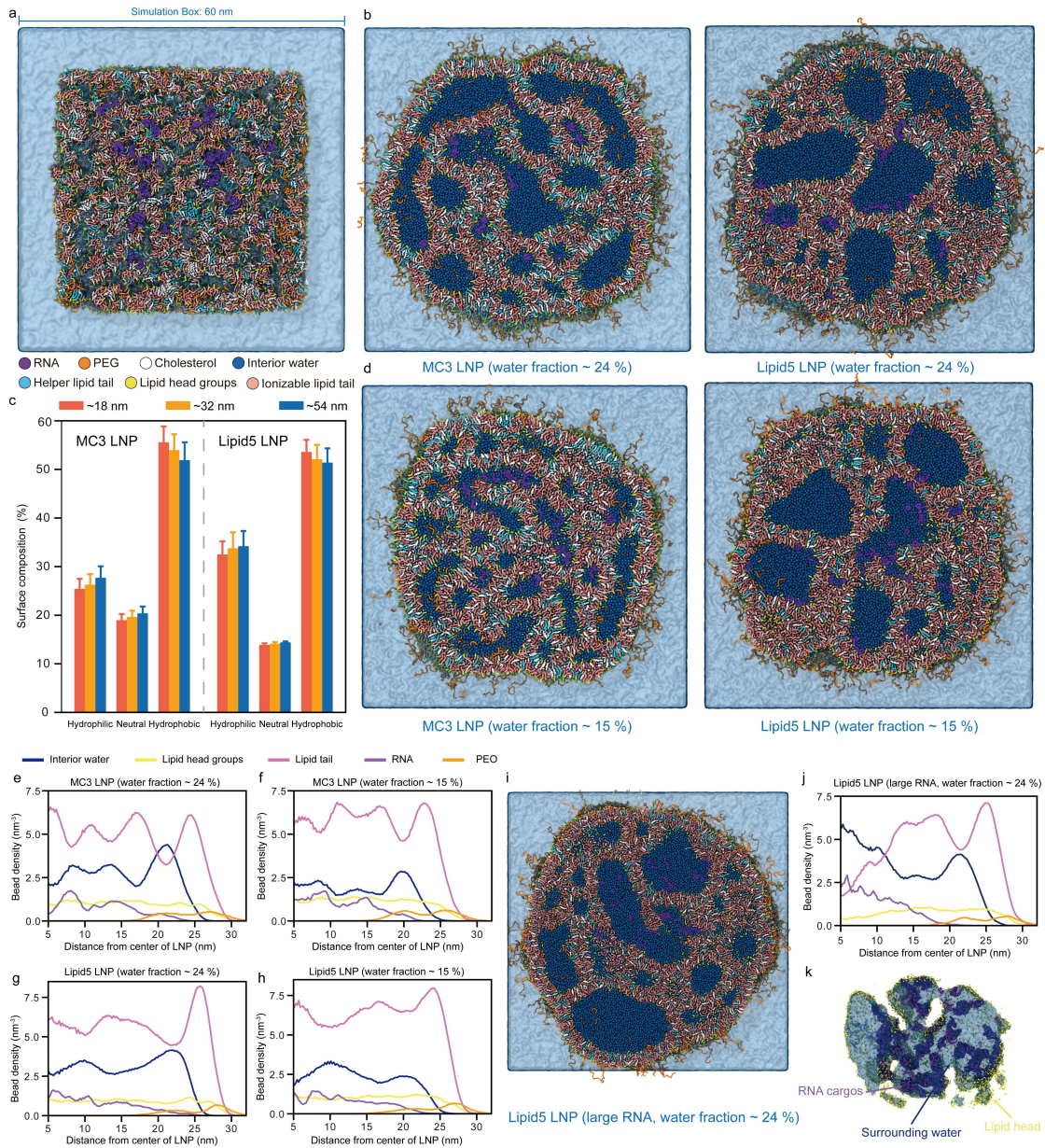

Fig. 2: Self-assembly of full-scale LNPs approaching experimental dimensions. a. Initial configuration for constructing full-size LNPs, where 56,840 lipids and 168 RNA strands were initially densely packed in a 47 nm box and solvated in a 60 nm water box. b. Cross-sectional views of the final self-assembled structures of MC3 and Lipid5 LNPs with ~24% interior water fraction. c. Comparison of surface hydrophilicity between large LNPs (this figure) and smaller LNPs (Fig. 1), shown as the relative fraction of hydrophilic, neutral, and hydrophobic regions. Values are expressed as mean ± s.d. d. Cross-sectional views of the final self-assembled structures of MC3 and Lipid5 LNPs with ~15% interior water fraction. e-h. Radial density profiles of MC3 LNPs at 24% (e) and 15% (f) water fractions, and Lipid5

LNPs at 24% (g) and 15% (h) water fractions, demonstrating the spatial distributions of RNA, PEG, lipid tails, lipid headgroups, interior water in LNPs. i. Cross-sectional snapshot of the Lipid5 LNP encapsulating larger RNA cargos. j. Radial density distribution of the Lipid5 LNP with large RNA cargos. k. Cross-sectional visualization of internal RNA aggregates (violet) localized within aqueous regions (blue) enclosed by lipid headgroups (yellow) in the Lipid5 LNP, highlighting interior RNA and water localization.

**Influence of LNP formulation ratios on structural and surface properties**

Beyond the lipid chemistry of ionizable lipids, the relative proportions of the lipid components in LNP formulations represent a critical design parameter that significantly influences the overall properties and functionality of LNPs[39-41]. To assess the effect of formulation composition, we evaluated three modified formulations that are ionizable lipid-enriched, cholesterol-enriched, and helper lipid-enriched, across three types of ionizable lipids: MC3, Lipid5, and ALC-0315. Representative cross-sectional views of the final self-assembled structures are shown in Fig. 3a-c, with empty LNP controls provided in Supplementary Fig. S7. Structural analyses were complemented by quantitative assessments of internal organization and surface hydrophilicity (Fig. 3d-i).

We observed that increasing the proportion of ionizable lipids had divergent effects on internal organization depending on the lipid type (Fig. 3d-f). In MC3-based LNPs, the number of internal reverse micelles decreased slightly with higher ionizable lipid content. By contrast, Lipid5 and ALC-0315 LNPs exhibited a substantial increase in reverse micelle formation under the same conditions, reflecting the stronger intrinsic propensity of these lipids to drive the formation of reverse micelles. However, higher ratios of the ionizable lipids led to a reduction in surface hydrophilicity and an increase in surface hydrophobicity across all three LNPs (Fig. 3g-i), potentially compromising colloidal stability and promoting aggregation.

Cholesterol enrichment elicited lipid-specific responses. In MC3-based LNPs, an increased cholesterol ratio modestly promoted the formation of internal reverse micelles. By contrast, in Lipid5 and ALC-0315 LNPs, the same enrichment resulted in a moderate reduction in micelle number. This divergence suggests that cholesterol selectively facilitates micellar organization when paired with ionizable lipids that exhibit limited intrinsic propensity for reverse micelle formation. Across all three LNP systems, elevated cholesterol levels also led to a subtle but consistent increase in surface hydrophilicity, indicative of a modest stabilizing effect on the LNP surface. Helper lipid-enriched formulations uniformly suppressed the formation of internal reverse micelles, regardless of ionizable lipid identity. This reduction in internal structuring may compromise the delivery efficiency of LNPs, particularly in RNA release. However, the helper lipid enriched formulation exhibited the highest surface hydrophilicity and lowest hydrophobicity among all conditions tested, characteristics associated with enhanced colloidal stability. Together, these results provide molecular-level insights that clarify previously observed experimental trends, notably the ability of phospholipids and cholesterol to enhance structural stability and reduce LNP size[12, 42].

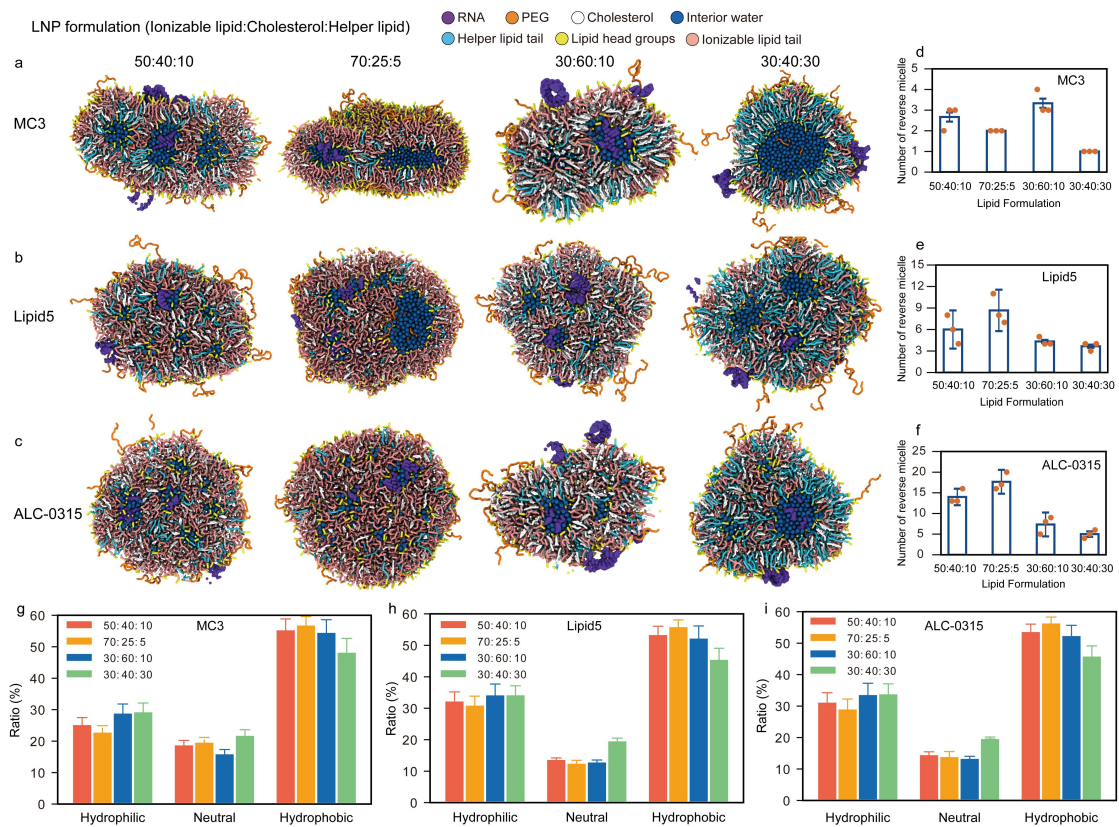

Fig. 3: Effect of LNP formulation composition on structure and surface properties. a-c. Final self-assembled structures of MC3 LNP (a), Lipid5 LNP (b), and ALC-0315

LNP (c) shown in cross-sectional view with varying formulations. LNP formulations are defined by the molar ratio of ionizable lipid: cholesterol: helper lipid, with their compositions examined: 50:40:10, 70:25:5, 30:60:10, and 30:40:30. d-f. Number of reverse micelles in MC3 (d), Lipid5 (e), and ALC-0315 (f) LNPs for different formulations. Data points were obtained from three individual simulations. g-i. Surface hydrophilicity of MC3 (g), Lipid5 (h), and ALC-0315 (i) LNPs across different formulations, quantified as the relative area composition of hydrophilic, neutral, and hydrophobic regions. All values are expressed as mean ± s.d. (n=3).

**Influence of pH transition on LNP structure**

Although LNPs are initially assembled under acidic conditions to promote RNA encapsulation, they are subsequently transferred to a neutral pH environment for storage and *in vivo* delivery[2]. This transition induces deprotonation of the ionizable lipids, converting them from a positively charged to a neutral state, which can markedly alter the overall architecture of the LNP. To probe this effect, we first examined MC3-based LNPs by selectively deprotonating only surface-localized MC3 molecules (Supplementary Fig. S8). We observed that deprotonated lipids migrated toward the LNP core, while helper lipids and protonated MC3 lipids in the interior LNP redistributed toward the surface and subsequently underwent deprotonation. This dynamic exchange, coupled with rapid solvent exchange between the LNP core and surrounding environment, suggested that pH-induced structural rearrangements were not confined to the surface but propagated throughout the LNP[43-45].

Motivated by this observation, we extended the analysis by globally deprotonating all ionizable lipids in LNPs previously assembled at pH 4, thereby capturing the full impact of the pH shift. Upon transition to pH 7.4, all LNPs became more spherical (Fig. 4a), likely due to decreased hydrophilicity of the ionizable lipid headgroups in their neutral state (Fig. 4b). This transformation was accompanied by a redistribution of lipid species: helper lipids and cholesterols became enriched at the surface, while ionizable lipids accumulated in the interior (Fig. 4c)[19, 45]. Notably, MC3 LNPs exhibited visible phase separation, with deprotonated MC3 lipids excluded from domains rich in helper lipids and cholesterol, which was not observed in Lipid5 or ALC-0315 LNPs, and likely attributable to the lower hydrophilicity of MC3 headgroups post-deprotonation. Across all LNPs, the number of internal reverse micelles decreased following deprotonation (Fig. 4d), indicating that the neutral form of ionizable lipids exhibits diminished capacity to stabilize highly curved aqueous domains. This trend is in strong agreement with experimental evidence demonstrating a pH-dependent reduction in internal mesophase curvature within LNPs under physiological conditions[46, 47].

We further evaluated these pH-induced effects in larger LNP systems generated in Fig. 2. MC3 LNP underwent pronounced structural rearrangements at pH 7.4, accompanied by partial RNA release from the interior (Fig. 4e). In contrast, Lipid5 LNPs maintained structural integrity and retained RNA cargo (Fig. 4f), likely due to the hydroxyl headgroup of Lipid5, which preserves hydrophilicity after deprotonation. Surface composition analysis confirmed a marked redistribution of lipid species in MC3 LNPs (Fig. 4g), with helper lipids and cholesterols migrating outward to stabilize the surface and deprotonated MC3 lipids concentrating in the core, a pattern that recapitulated experimental observations of phospholipid-enriched surfaces at physiological pH[19, 36, 48]. Furthermore, hydration analysis of lipids revealed that more than half of the deprotonated MC3 lipids were dehydrated in the interior and clustered with cholesterols to form a dense lipid phase (Fig. 4h). This result mirrors previous Cyro-TEM studies showing that neutralized ionizable lipids partition into the LNP interior to minimize hydration[45, 49]. Lipid5 LNPs also displayed similar, but less pronounced, formation of dehydrated lipid domains (visualization of the distribution of dehydrated lipids in MC3 and Lipid 5 LNPs was shown in Supplementary Fig. S9). Finally, the radial density profiles at pH 7.4 (Fig. 4i-j) illustrated a decrease in the layered internal organization, particularly in MC3 LNPs, reinforcing the notion that pH transitions compromise both surface and core structures. This effect also was observed in LNPs containing large RNAs (Supplementary Fig. S10). Taking together, these results suggest that deprotonation leads to the collapse of inverse micellar architecture, whereby aqueous compartments merge and neutral lipids coalesce with cholesterol into an internal lipid-rich phase, especially in high water-content systems such as mRNA-loaded LNPs[38, 50].

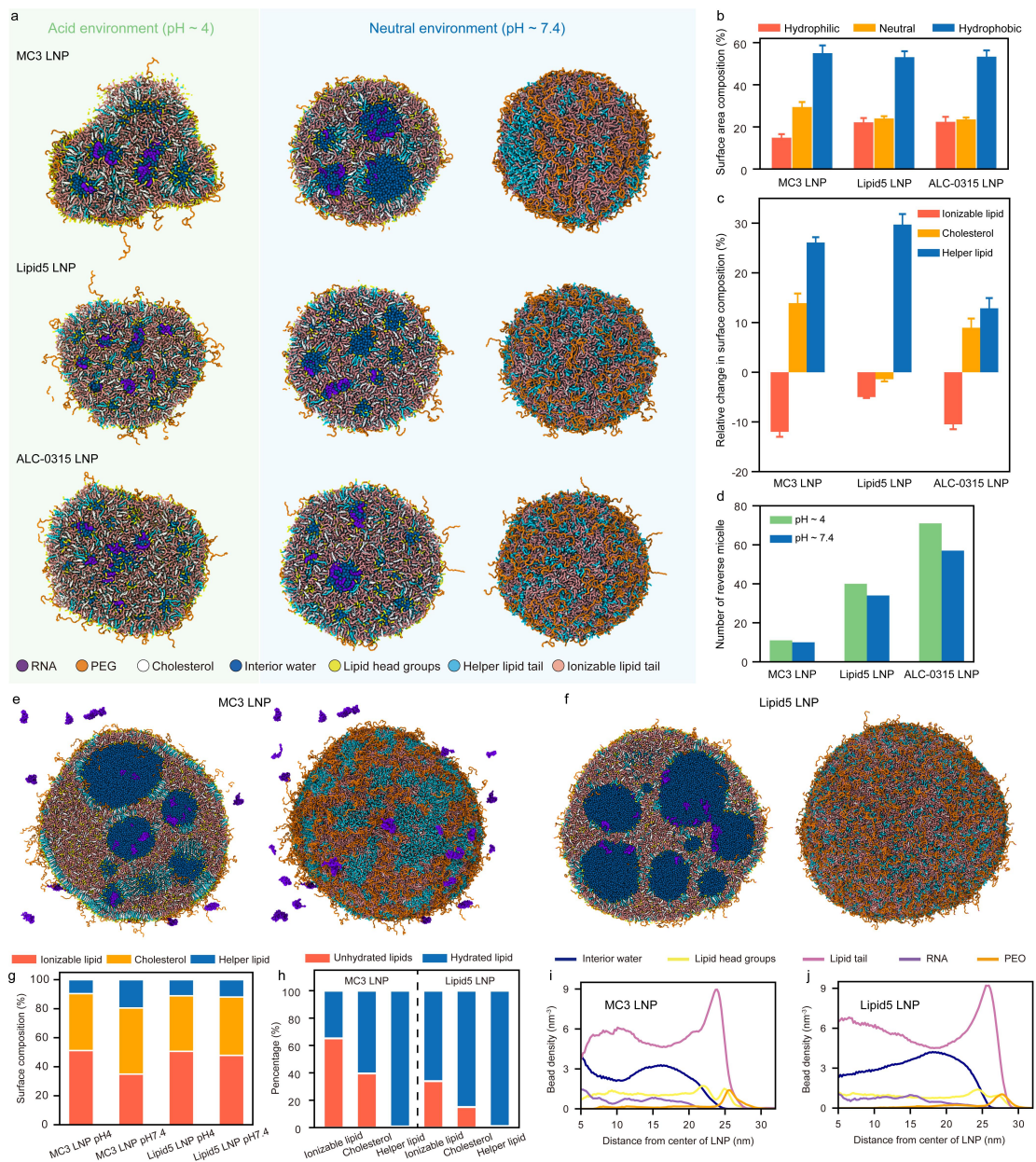

Fig. 4: Effects of pH transition on LNP structure and surface properties. a. Cross-sectional views of self-assembled MC3, Lipid5, and ALC-0315 LNPs under acidic conditions (~pH 4) and their structural transformation in a neutral environment (~pH 7.4) shown in cross-sectional view and overall view (lipid headgroups are shown in the same color as their corresponding tails to better visualize lipid aggregation). b. Surface hydrophilicity of MC3, Lipid5, and ALC-0315 LNPs at pH 7.4, represented as the relative composition of hydrophilic, neutral, and hydrophobic regions. Data are presented as mean ± s.d. (n=3) c. Relative change in surface composition between pH 4 and pH 7.4, expressed as a percentage change in molar

ratios of ionizable lipids, cholesterol, and helper lipids. Data are presented as mean ± s.d. (n=3) d. Number of reverse micelles in MC3, Lipid5, and ALC-0315 LNPs at pH 4 and pH 7.4, showing a reduction in reverse micelles upon transitioning to neutral pH. e-f. Cross-sectional and overall views of large MC3 (e) and Lipid5 (f) LNPs (~54 nm in diameter) at pH 7.4, illustrating the structural transformation in both interiors and surface of the LNPs. g. Surface composition of large MC3 and Lipid5 LNPs at pH 4 and pH 7.4, showing differences in ionizable lipid, cholesterol, and helper lipid distribution. h. Proportion of hydrated versus dehydrated lipids in the interior of MC3 and Lipid5 LNPs at pH 7.4, categorized by lipid type. i-j. Radial density distribution of different molecular components in MC3 (i) and Lipid5 (j) LNPs at pH 7.4, depicting the spatial organization of RNA, PEG-lipids, lipid tails, lipid head groups, and interior water.

**Rational design of novel ionizable lipids and optimized LNP formulation**

Building on the mechanistic insights derived from our CGMD simulations, we demonstrate how this strategy can inform the rational design of novel ionizable lipids and their optimized LNP formulations, in conjunction with experimental validation. Fig. 5a presents different views of self-assembled LNPs composed of four different novel ionizable lipids, each formulation under a standard molar ratio. LNPs with low encapsulation efficiency (EE) exhibited disordered internal organizations, characterized by poorly packed lipids and diffuse distribution of aqueous regions, indicating their poor capacity to form reverse micelles and encapsulate RNA effectively. This disordered morphology likely results from the insufficient hydrophobic tail lengths of the ionizable lipids, which are unable to drive proper phase segregation and reverse micelle formation. By contrast, LNPs with high EE featured well-defined reverse micelles, wherein internal aqueous regions were fully enclosed by lipid headgroups, correlating with improved RNA loading capacity.

To extend these insights toward formulation optimization, we investigated the relationship between ionizable lipid ratio and LNP surface properties using Lipid L5 as a model compound (Fig. 5b). Lipid L5 adopts a classic cone-shaped geometry, with small hydrophilic headgroups and bulky hydrophobic tails. Increasing its molar fraction in the formulation elevated surface hydrophobicity (Fig. 5c) and resulted in a pronounced increase in hydrodynamic diameter (Fig. 5b), suggesting that excessive hydrophobicity destabilizes the LNP interface and drives LNP growth into larger surface area to restore stability. Conversely, reducing the ratio of Lipid L5 increased surface hydrophilicity and yielded a smaller LNP size. Additionally, PEG chains were found to extend further from the LNP surface in formulations with reduced Lipid L5 content (Fig. 5d), which enhances steric repulsion and suppresses aggregations[42].

To assess the impact of ionizable lipid headgroup chemistry, we analyzed LNPs containing Lipid L6 (Fig. 5e), a structurally distinct ionizable lipid featuring larger hydrophilic headgroups. Compared to Lipid L5 LNPs, Lipid L6-based LNPs exhibited substantially higher surface hydrophilicity (Fig. 5f). Notably, increasing the proportion of Lipid L6 in the formulation did not appreciably affect surface hydrophilicity or LNP size (Fig. 5e), different from the behavior observed with Lipid L5. Together, these results offer a framework for guiding formulation strategies based on ionizable lipid structure. For cone-shaped lipids with small hydrophilic headgroups and bulky hydrophobic tails, increasing their formulation ratio elevates surface hydrophobicity, which causes the increase of LNP with poor stability. Conversely, ionizable lipids with inherently large hydrophilic headgroups can be used at higher ratios of ionizable lipids without compromising stability.

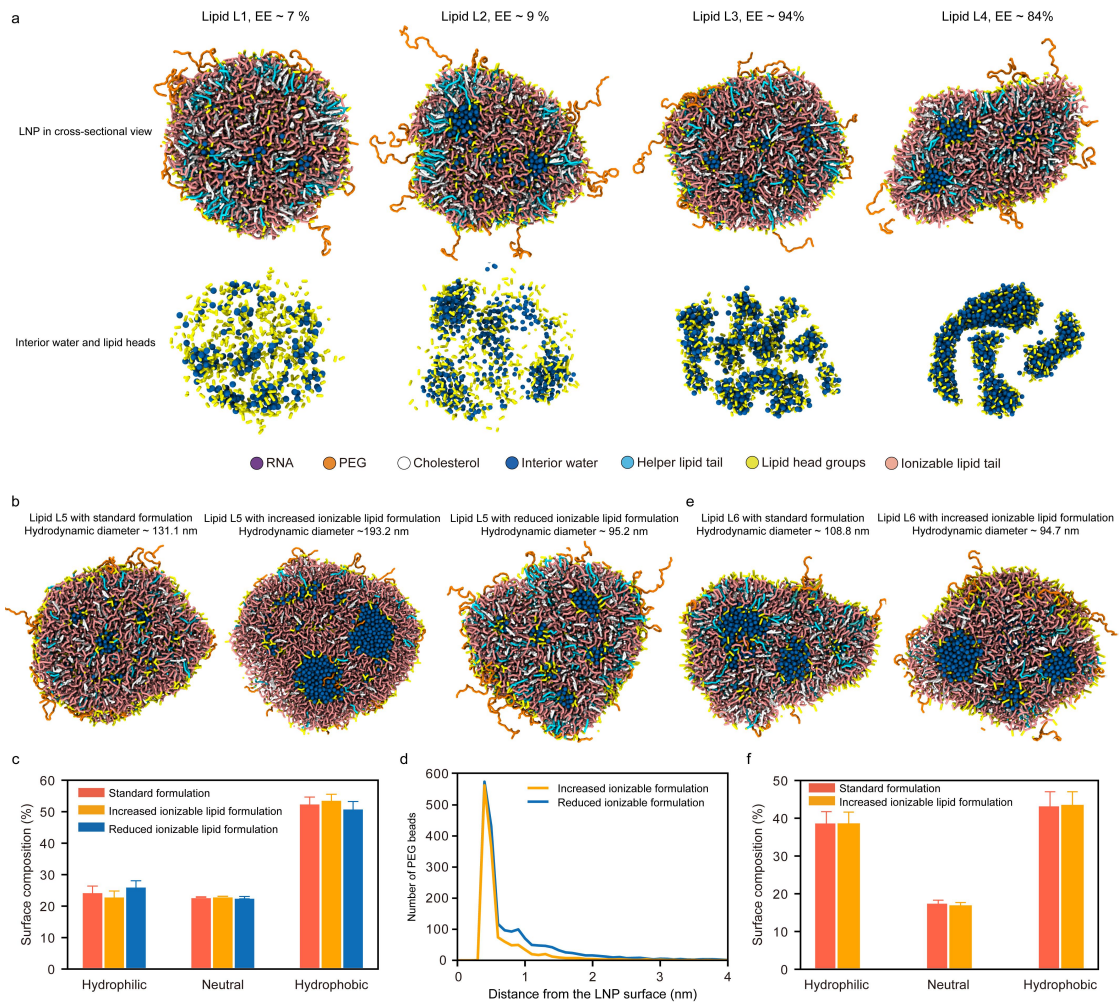

Fig. 5: Rational design of novel ionizable lipids and optimized LNP formulation. a. Cross-sectional views of self-assembled LNPs composed of four distinct novel ionizable lipids under standard formulation ratios. Bottom panels display only the interior water and lipid headgroups, revealing a strong correlation between EE and the presence of well-defined reverse micelle internal structures. b. Cross-sectional views of LNPs containing Lipid L5 at varying formulation ratios, illustrating that hydrodynamic diameter is highly sensitive to ionizable lipid content. The standard, ionizable lipid-enriched, and ionizable lipid-reduced formulations correspond to molar ratios of ionizable lipid: cholesterol: helper lipid: PEG-lipid of 50:38.5:10:1.5, 70:24:4:2, and 35:46.5:16:2.5, respectively. c. Surface composition profiles of Lipid L5 LNPs, quantifying the relative contributions of hydrophilic, neutral, and hydrophobic regions at the LNP surface across different formulations. Data are presented as mean ± s.d. (n=3) d. Radial distribution of PEG chains on Lipid L5 LNPs,

illustrating greater extension from the surface at lower ionizable lipid ratios in formulation. e. Cross-sectional views of LNPs containing Lipid L6, an ionizable lipid with larger hydrophilic headgroups. LNPs show minimal changes in size with varying formulation ratios compared to Lipid L5. f. Surface composition profiles of Lipid L6 LNPs, indicating a consistently high surface hydrophilicity independent of formulation ratio. Data are presented as mean ± s.d. (n=3).

**Conclusion**

We present a CGMD-based framework that reveals the molecular-scale mechanisms driving the self-assembly and structural organization of LNPs. By resolving the impact of ionizable lipid chemistry, formulation ratios and pH transitions, we identify key molecular features that govern the internal architecture and surface hydrophilicity of LNPs. We further demonstrate how we use this computational framework to pre-screen novel lipid candidates and choose LNP formulation ratios by examining the internal structure and surface hydrophobicity of LNPs in simulations. These findings advance the mechanistic understanding of the LNP structure and the intricate interactions within the LNP components at the molecular level and highlight its importance in rational design of novel LNPs.

**Methods**

**Modeling and parameterization for CGMD simulations**

The CG structure and force field parameters of ionizable lipids were derived through a multistep procedure. First, by providing the SMILES strings of the ionizable lipids, we used the CGenFF toolbox to obtain the all-atom (AA) structure and corresponding parameters of ionizable lipids under the CHARMM force field[51, 52]. Next, AA molecular dynamics (AAMD) simulations were performed on lipid bilayers composed solely of protonated ionizable lipids. The resulting AAMD trajectories were then mapped to CG trajectories using the PyCGTOOL.py script[53], following the Martini 2.0 force field convention[26], where four heavy atoms are grouped into a single CG bead. The bond and angle parameters of the CG ionizable lipids were determined by fitting the bond and angle distributions from AAMD trajectories. Supplementary Fig. S11-S16 presents the CG structures of MC3, Lipid5, and ALC-0315, along with their fitting bond and angle distributions compared to their AAMD counterparts. The CG RNA structures and parameters were obtained using the martinize-nucleotide.py script[54]. The details of simulation settings were provided in Supplementary Information.

**LNP Synthesis**

The novel ionizable lipids were prepared by classical chemical reactions and purified by chromatography techniques at Pharmaron (Beijing, China). The identity and purity of these lipids were confirmed by liquid chromatography-mass spectrometry and H NMR. While all other lipids (DSPC lipids and PEG lipids) were purchased from Avanti Polar Lipids. Cholesterol was purchased from Sigma Aldrich. LNPs were synthesized by mixing an organic phase, containing all lipid components, with an aqueous phase containing the nucleic acid payload (e.g., mRNA). The organic phase was prepared by dissolving all the lipids in pure ethanol, while the aqueous phase consisted of a citrate buffer at pH 4. Mixing was performed using a microfluidic chip at an organic-to-aqueous flow rate ratio of 1:3. Immediately following mixing, the solution underwent a dilution step, ethanol removal, buffer exchange, and concentration using centrifugal ultrafiltration. The final mRNA concentration in the LNP formulation was adjusted based on the measured encapsulation efficiency (EE).

**EE determination**

EE was determined by quantifying both the total mRNA content (encapsulated and unencapsulated) and the unencapsulated mRNA remaining in solution. The Quant-iT RiboGreen RNA Assay (Invitrogen) was used for fluorescence-based quantification. To measure total mRNA, samples were treated with 0.5% (w/w) Triton X-100 to disrupt the LNPs and release encapsulated mRNA, followed by the addition of RiboGreen reagent. Fluorescence intensity was measured using a microplate reader and compared to an RNA standard curve. To measure unencapsulated mRNA, the same procedure was followed without Triton X-100, so only free RNA in the solution

was detected. EE was calculated using the following equation:

$$Encapsulation\ Efficiency\ (EE, \%) = (\frac{Total\ mRNA - free\ mRNA}{Total\ mRNA}) \times 100$$

**LNP Size Measurement**

LNP size and size distribution were measured using Dynamic Light Scattering (DLS) with a Malvern Zetasizer at 25°C. Each sample was analysed in triplicate. Prior to measurement, samples were diluted in an appropriate buffer (e.g., DPBS or Tris buffer) to achieve optimal particle concentration for DLS. Instrument settings such as refractive index and dispersant viscosity were configured accordingly. The Z-average diameter, representing the intensity-weighted mean hydrodynamic size, was reported as the primary size metric. Additionally, the polydispersity index (PDI) was recorded to assess the uniformity of the particle size distribution.

**Author Contributions**

X.B. and K.W. conceived and designed the research. X.B. performed the CGMD simulations and analysed the data. Y.L. conducted the experimental validation. X.B., T.Y., and K.L. developed the parameterization protocols for the CG models of ionizable lipids. C.Y. and F.S. contributed to the design of novel ionizable lipids. X.B. drafted the manuscript, and all authors reviewed, discussed, and provided input on the final version.

**Competing interests**

All authors are current employees of METiS Pharmaceuticals.

**Acknowledgement**

This project was supported by METiS Pharmaceuticals.

# Supplementary Information for

# Unraveling the Molecular Structure of Lipid Nanoparticles through *in-silico* Self-Assembly for Rational Delivery Design

Xuan Bai*, Yu Lu, Tianhao Yu, Kangjie Lv, Cai Yao, Feng Shi, Andong Liu, Kai Wang*, Wenshou Wang, Chris Lai

METiS Technology (Hangzhou) Co., Ltd, Hangzhou, China, 310000

Emails: xbai@metispharma.com, baixuan@zju.edu.cn; kwang@metispharma.com.

## Supplementary Methods

### Simulation Settings

The molecular dynamics (MD) simulations were conducted using the GROMACS package version 2022[1]. The temperature was maintained at 310 K with a coupling constant of 1.0 ps using v-rescale thermostat. Periodic boundary conditions were uniformly applied in the x, y, and z dimensions. Pressure conditions were controlled by Parrinello-Rahman method in all three dimensions, with a coupling constant of 12.0 ps and compressibility of $3*10^{-4}$. The Lennard-Jones potential was assigned a cutoff of 1.1 nm, which was smoothly shifted to zero to 1.1 nm to minimize cutoff noise. Electrostatic interactions were handled by reaction-field with cutoff of 1.1 nm. Each simulation was performed with a time step of 10 fs, and the neighbor list was updated every 20 steps. Snapshots were generated using the VMD package[2]. Details of the setup of systems can refer to Figure 1 and Supplementary Figure S1, S2 and S6.

## Supplementary Figures

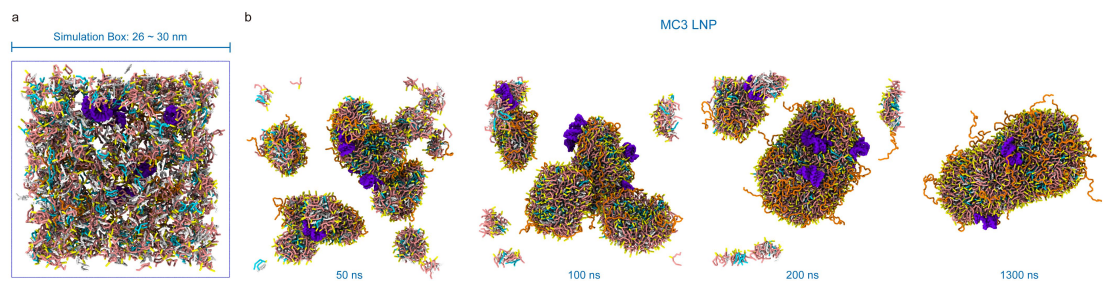

Supplementary Figure S1. Initial setup and time-resolved visualization of LNP

self-assembly in a small simulation system, using MC3 as an example. a. The simulation system consists of 1,000 ionizable lipids, 800 cholesterol molecules, 200 helper lipids, 30 PEG-lipids, and 6 RNA strands. All components were initially packed into a compact box (25-29 nm) based on the dimensions of the ionizable lipid and subsequently solvated in a larger box (26-30 nm). b. Snapshots of the MC3 LNP self-assembly process over time. Dispersed lipid and RNA components first form micellar aggregates, which progressively coalesce into a single mature LNP. The complete self-assembly process occurs over a timescale from several hundred to approximately 2,000 nanoseconds.

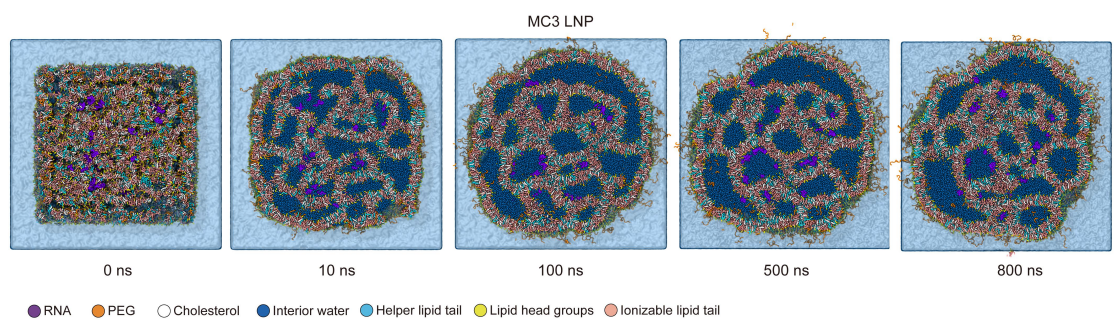

Supplementary Figure S2. Initial setup and time-resolved visualization of the construction of a near-physiological-size MC3-based LNP. The initial dense packing of LNP components was generated through a multistep process using the Packmol package. A central core composed of ionizable lipids, cholesterol, helper lipids, and RNA molecules was first packed into a 35 nm box. This core was then surrounded by six additional blocks containing ionizable lipids, cholesterol, helper lipids, and PEG-lipids, resulting in a fully constructed initial system within a 45 nm simulation box. The time-resolved snapshots show the rapid formation of the LNP, which reaches a stable, organized structure within a few hundred nanoseconds.

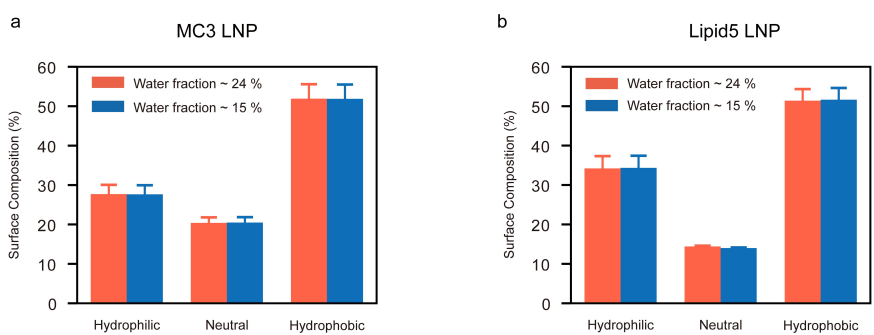

Supplementary Figure S3. Surface hydrophilicity of MC3 and Lipid5 LNPs with varying internal water content. Data are presented as means ± SD (n = 3). (a) MC3-based LNPs and (b) Lipid5-based LNPs exhibit comparable surface hydrophilicity profiles at two internal water levels (~24% and ~15%). Surface composition is categorized into hydrophilic, neutral, and hydrophobic regions. This shows minimal changes in hydrophilic and hydrophobic surface proportions between different hydration levels, indicating that surface hydrophilicity remains largely unaffected by the internal water content.

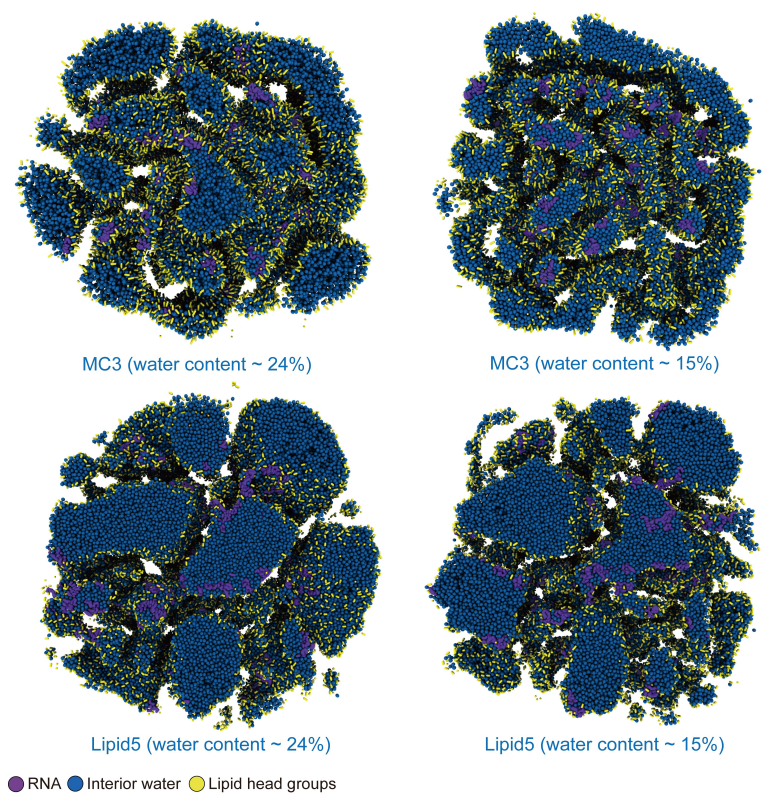

Supplementary Figure S4. Internal structural comparison of MC3 and Lipid5 LNPs with different internal water content. (a) MC3-based LNP and (b) Lipid5-based LNP visualized with two levels of internal hydration (~24% and ~15%). Only the interior water molecules (blue), encapsulated RNA strands (purple), and lipid head groups (yellow) surrounding the water compartments are shown to highlight structural organization in response to water content variation.

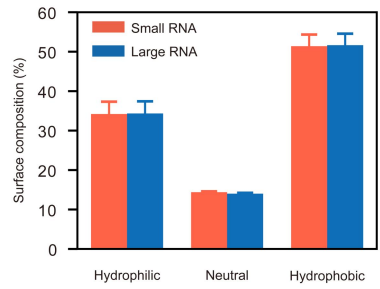

Supplementary Figure S5. Surface hydrophilicity of Lipid5 LNPs encapsulating different RNA cargos.

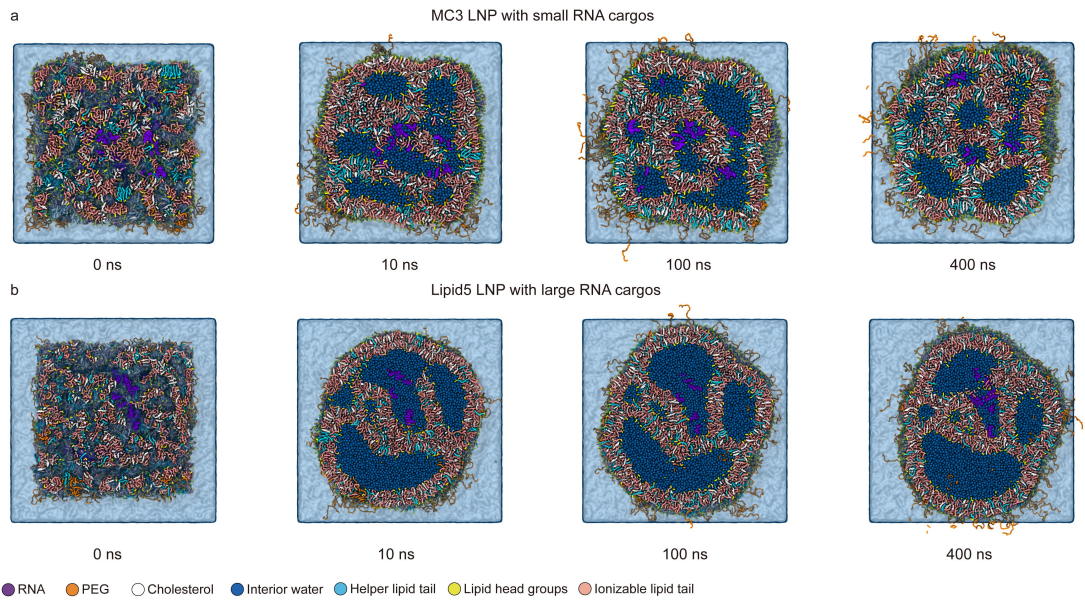

Supplementary Figure S6. Initial setup and time-resolved visualization of the fast construction of an entire LNP with size of about 30 nm.

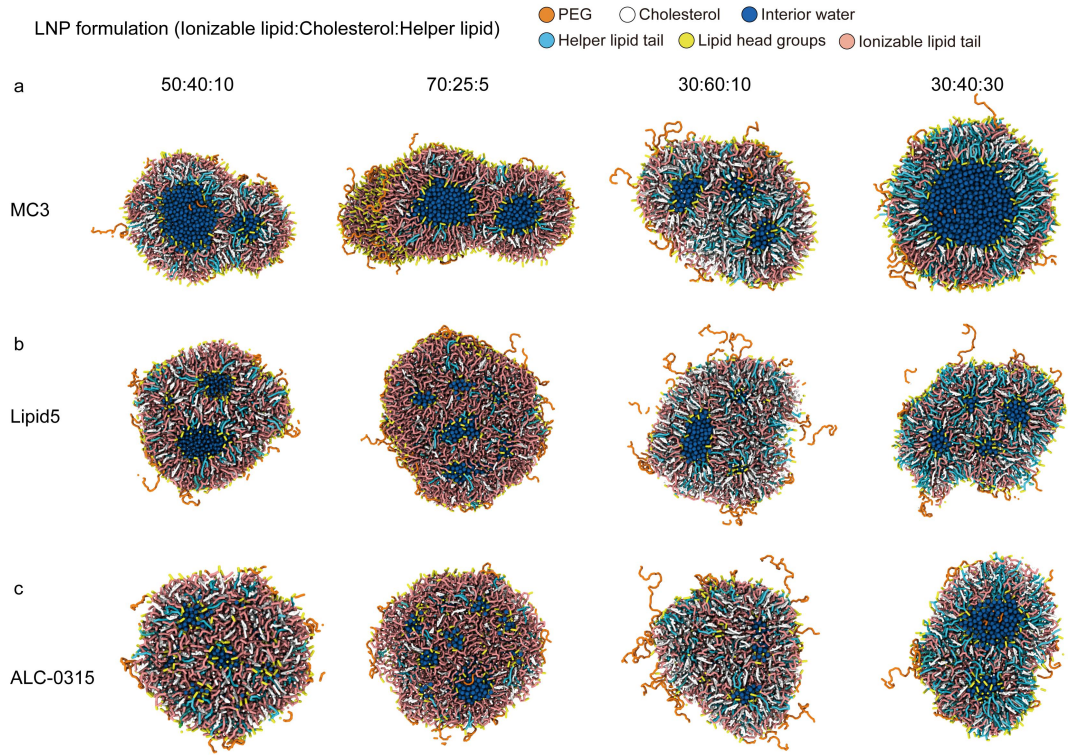

Supplementary Figure S7. Final self-assembled structures of MC3, Lipid5, and ALC-0315 LNPs with varying formulations in the absence of RNA cargo. Cross-sectional views of LNPs composed of (a) MC3, (b) Lipid5, and (c) ALC-0315 ionizable lipids, assembled under four distinct formulation ratios (ionizable lipid: cholesterol: helper lipid): 50:40:10, 70:25:5, 30:60:10, and 30:40:30. These structures reveal lipid-specific internal organization responses to formulation variation.

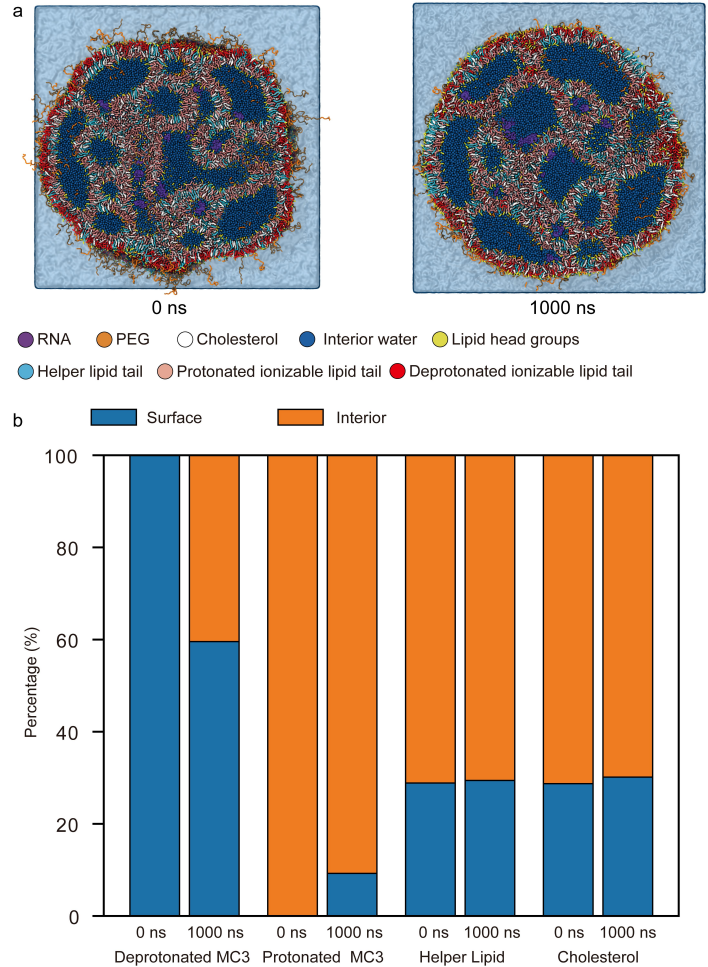

Supplementary Figure S8. Lipid redistribution within LNPs following surface-localized deprotonation of ionizable lipids. (a) Cross-sectional views of an MC3-based LNP at 0 ns, in which only surface-localized MC3 lipids are initially deprotonated (shown in red), and at 1000 ns, showing the resulting lipid rearrangement. Over the simulation, deprotonated MC3 lipids migrate toward the interior, while protonated MC3 lipids and helper lipids redistribute to the surface. (b) Quantitative analysis of surface versus interior localization for deprotonated MC3, protonated MC3, helper lipids, and cholesterol at 0 ns and 1000 ns, highlighting the lipid exchange dynamics and compositional shift induced by deprotonation.

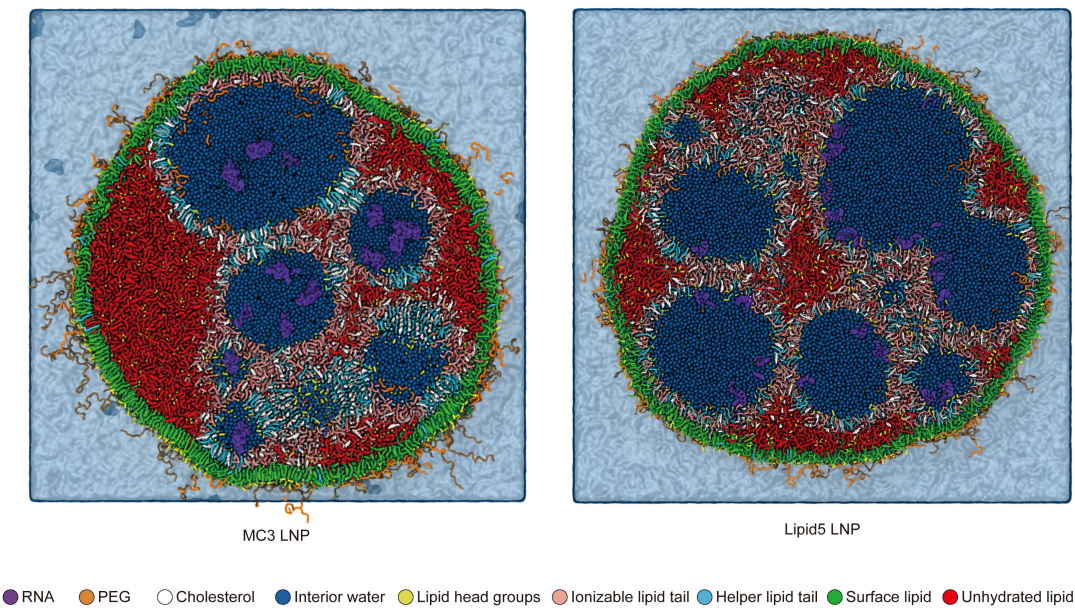

Supplementary Figure S9. Visualization of the distribution of dehydrated lipids in MC3 and Lipid5 LNPs. Cross-sectional views of MC3 (left) and Lipid5 (right) LNPs highlight regions where deprotonated ionizable lipids in the interior form dehydrated domains (red) into pure lipid phases. This behavior reflects the tendency of neutralized ionizable lipids to avoid aqueous environments, contributing to internal structural reorganization.

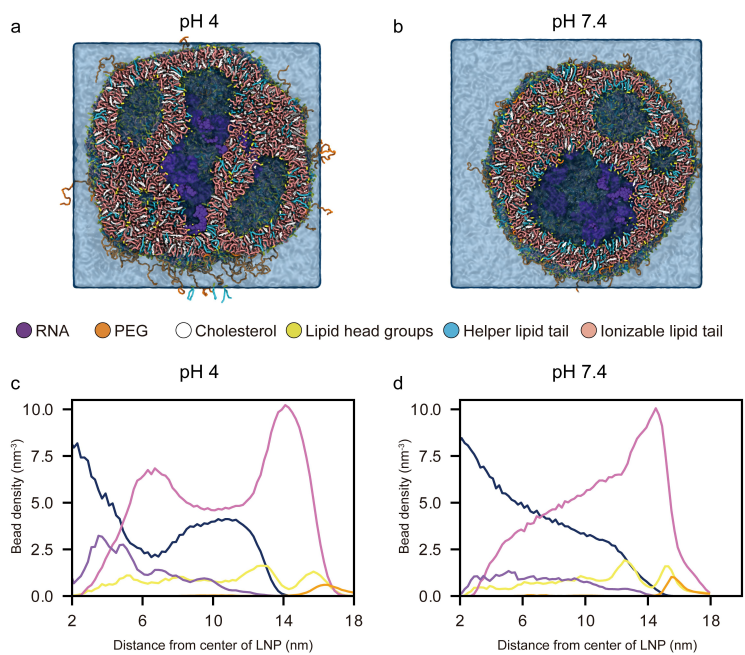

Supplementary Figure S10. Structural reorganization of Lipid5 LNPs encapsulating

large RNA cargos during pH transition. Cross-sectional views of Lipid5 LNPs at pH 4 (a) and pH 7.4 (b). Quantitative bead density profiles as a function of distance from the LNP center reveal reduced internal layering upon transition from pH 4 (c) to pH 7.4 (d).

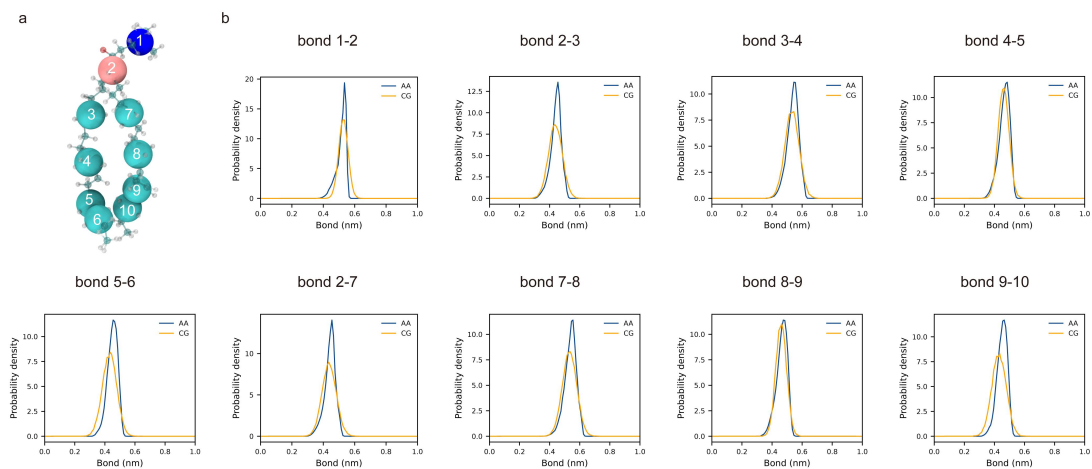

Supplementary Figure S11. CG model and bond parameter validation for MC3 in its charged state. a. Schematic illustration of the CG mapping scheme for the ionizable lipid MC3, with bead indices used to define the bonded interactions. The CG bead types assigned are Q0 (1), Na (2), C1 (3, 4, 7, 8), and C4 (5, 6, 9, 10). b. Probability density distributions of bond lengths obtained from AA and CG simulation trajectories for the specified bonds. The close agreement between AA and CG profiles confirms the accuracy of CG bond parameterization in reproducing molecular geometry.

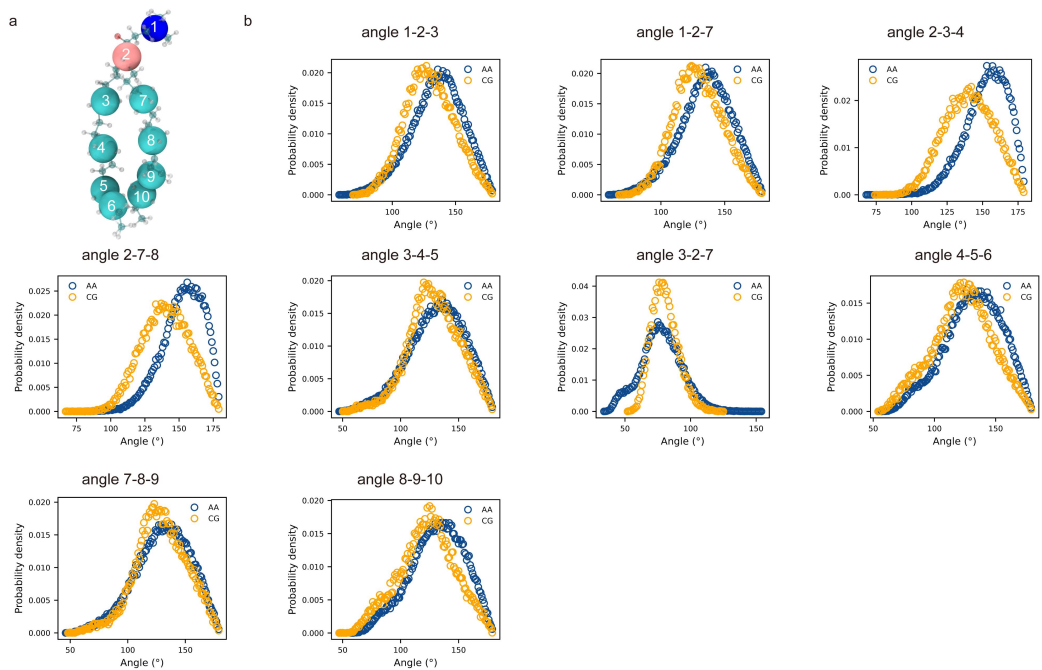

Supplementary Figure S12. CG model and angle parameter validation for MC3 in its charged state. a. Schematic illustration bead indices of MC3 lipid used to define angles. b. Probability density distributions of angles obtained from AA and CG simulation trajectories. The overall agreement between AA and CG profiles supports the accuracy of the CG parameterization in capturing the angular geometry of MC3.

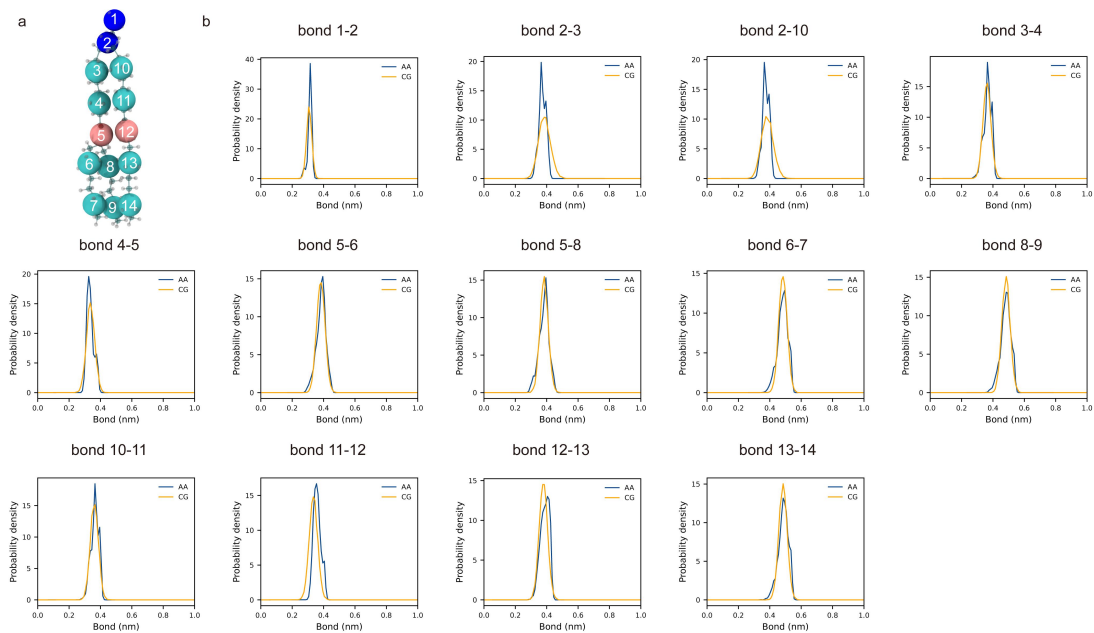

Supplementary Figure S13. CG model and bond parameter validation for Lipid5 in its charged state. a. Schematic illustration of the CG mapping scheme for the ionizable

lipid Lipid5, with bead indices used to define the bonded interactions. The CG bead types assigned are SP2 (1), Q0 (2), C2 (3, 4, 10, 11), Na (5, 12), and C1 (6, 7, 8, 9, 13, 14). b. Probability density distributions of bond lengths obtained from AA and CG simulation trajectories for the specified bonds.

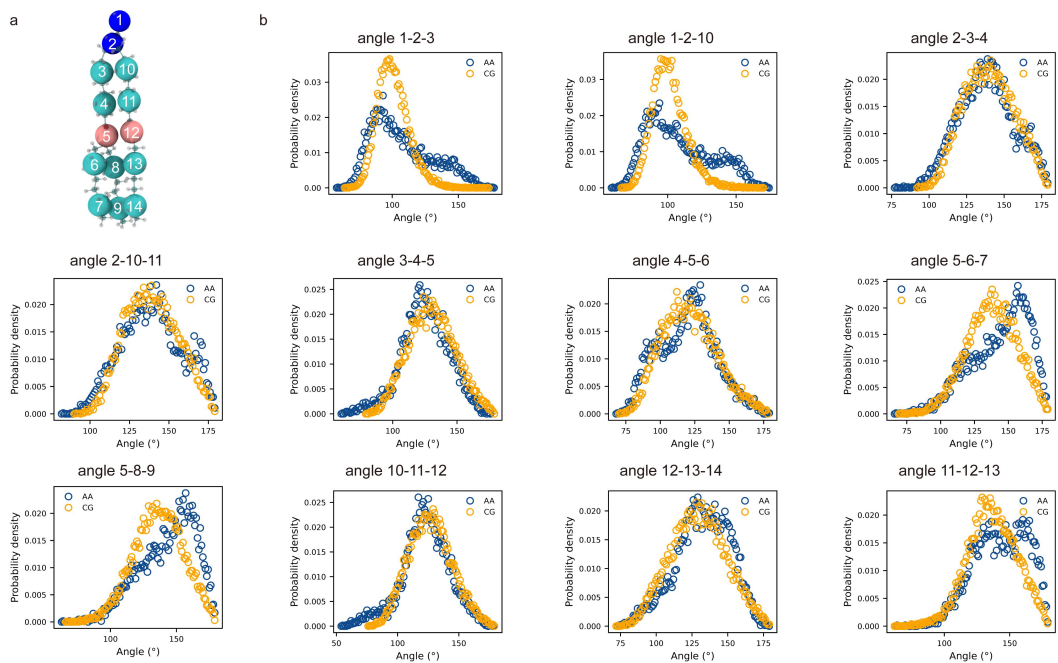

Supplementary Figure S14. CG model and angle parameter validation for Lipid5 in its charged state. a. Schematic illustration bead indices of Lipid5 used to define angles. b. Probability density distributions of angles obtained from AA and CG simulation trajectories.

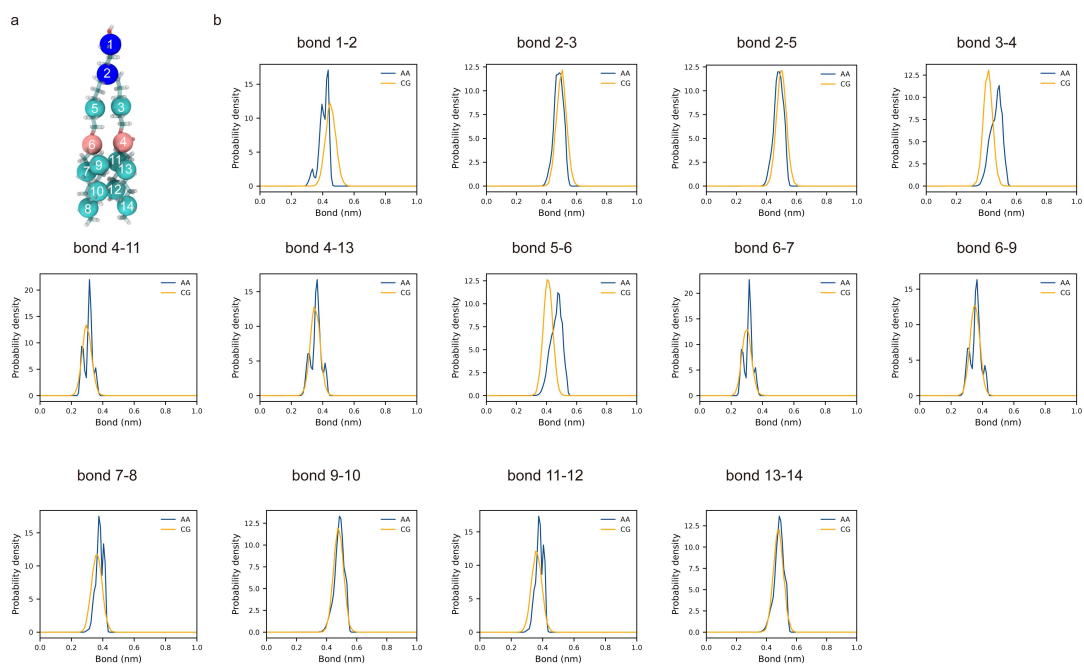

Supplementary Figure S15. CG model and bond parameter validation for ALC-0315 in its charged state. a. Schematic illustration of the CG mapping scheme for the ionizable lipid ALC-0315, with bead indices used to define the bonded interactions. The CG bead types assigned are P1 (1), Q0 (2), C1 (3, 5, 7, 8, 9, 10, 11, 12, 13, 14), and Na (4, 6). b. Probability density distributions of bond lengths obtained from AA and CG simulation trajectories for the specified bonds. The close agreement between AA and CG profiles confirms the accuracy of CG bond parameterization in reproducing molecular geometry.

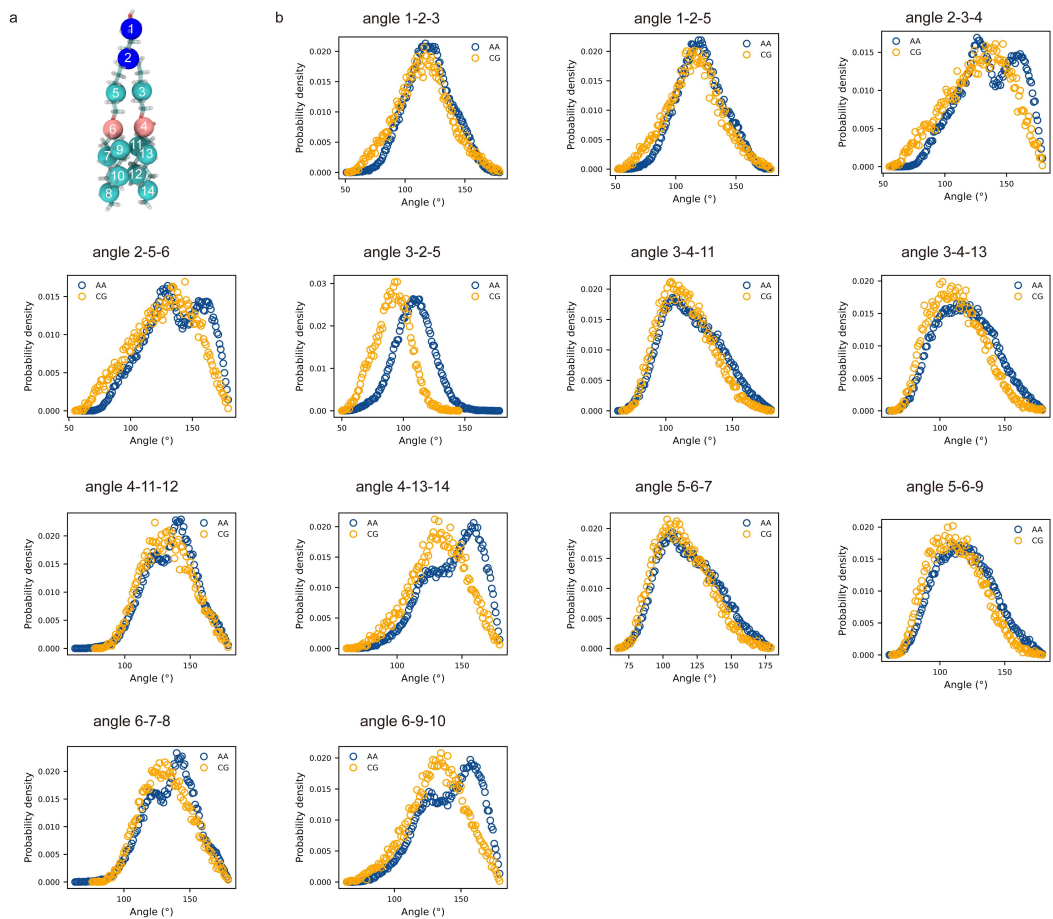

Supplementary Figure S16. CG model and angle parameter validation for ALC-0315 in its charged state. a. Schematic illustration bead indices of ALC-0315 used to define angles. b. Probability density distributions of angles obtained from AA and CG simulation trajectories.